\documentclass[12pt]{article}   	
\usepackage{geometry}                		
\usepackage{graphicx}				
\usepackage{amssymb}
\usepackage{amsmath}
\usepackage{epstopdf}
\usepackage{natbib}
\usepackage{hyperref}
\usepackage{setspace}
\usepackage{afterpage}
\usepackage{authblk}
\usepackage{pdflscape}
\usepackage{threeparttable}
\hypersetup{colorlinks=true, linkcolor=blue, citecolor=blue,urlcolor=blue}
\usepackage[all]{hypcap}
\usepackage{lineno}
\usepackage{authblk}
\usepackage[labelfont=bf]{caption}
\usepackage{color}
\usepackage{comment}

\usepackage[normalem]{ulem}

\def\eqref#1{(\ref{eq:#1})}

\newcommand{\tobs}{t_{\mathrm{obs}}}
\newcommand{\tlife}{t_{\mathrm{life}}}

\title{On the Rate of Abiogenesis \\from a Bayesian Informatics Perspective}
\author[1,*]{Jingjing Chen}
\author[1]{David Kipping}
\date{}							

\affil[1]{Department of Astronomy, Columbia University, 550 W 120th St., New York NY 10027}
\affil[*]{Corresponding author. E-mail: jc4132@columbia.edu.}

\begin{document}
\maketitle





\begin{abstract}

Life appears to have emerged relatively quickly on the Earth, a fact
sometimes used to justify a high rate of spontaneous abiogenesis ($\lambda$)
among Earth-like worlds. Conditioned upon a single datum - the time of earliest
evidence for life ($t_{\mathrm{obs}}$) - previous Bayesian formalisms for the
posterior distribution of $\lambda$ have demonstrated how inferences are highly
sensitive to the priors. Rather than attempt to infer the true
$\lambda$ posterior, we here compute the relative change to $\lambda$ when new
experimental/observational evidence is introduced. By simulating posterior
distributions and resulting entropic information gains, we compare three
experimental pressures on $\lambda$: 1) evidence for an earlier start to life;
$t_{\mathrm{obs}}$; 2) constraints on spontaneous abiogenesis from the lab; and
3) an exoplanet survey for biosignatures. First, we find that experiments
1 and 2 can only yield lower limits on $\lambda$, unlike 3. Second,
evidence for an earlier start to life can yield negligible information on
$\lambda$ if $t_{\mathrm{obs}} \ll \lambda_{\mathrm{max}}^{-1}$.
Vice versa, experiment 2 is uninformative when $\lambda_{\mathrm{max}} \gg
t_{\mathrm{obs}}^{-1}$. Whilst experiment 3 appears the most direct means of
measuring $\lambda$, we highlight that early starts inform us of
the conditions of abiogenesis, and that lab experiments could succeed
in building new life. Together then, the three experiments are complimentary
and we encourage activity in all to solve this grand challenge.

\end{abstract}

\clearpage

\section{Introduction}
\label{sec:intro}

The number of communicative civilizations in our galaxy is often framed in
terms of the famous Drake Equation, described by seven parameters which can be
considered to guide the observational inputs needed to make an inference. When
first proposed by Frank Drake over a half a century ago, all of the terms were
essentially unknown \citep{drake:1965}, but today observational constraints
exist for the first three terms; the rate of star formation
\citep{robitaille:2010}, the fraction of stars with planets
\citep{cassan:2012}, and the average number of habitable planets per star
\citep{dfm:2014,dressing:2015}. It is therefore timely to consider what type
of observations might enable us to best constrain the next term, $f_l$, the
fraction of habitable planets which develop life.

At the time of writing, interstellar panspermia is considered to be a
unlikely pathway for planets to developing life \citep{wessen:2010},
which leaves the most likely pathway as being abiogenesis: the spontaneous
process by which inert matter becomes living.

The rate of abiogenesis, $\lambda$, is surely functionally dependent upon the
very specific geophysical, chemical and environmental conditions of a given
world \citep{scharf:2016}. For example, it would seem manifestly wrong to
assume that frigid Pluto should have the same rate of abiogenesis events as the
early Earth. Consequently, one might reasonably ask whether knowledge of the
Earth's rate of abiogenesis has any useful merit for predicting the prevalence
of life elsewhere? 
Whilst this is certainly a reasonable and valid concern, we
are frankly left with no choice but to operate under the assumption that the
Earth and Earth-like exoplanets belong to the same distribution.

This concern can
at least be somewhat mitigated against by assuming that the Earth's rate of
abiogenesis is a representative sample drawn for a particular subset of planet
types - namely ``Earth-like'' planets. The precise definition of what
constitutes a member of this subset is an in-depth discussion in of itself
\citep{tasker:2017}, but let us continue under the assumption that such a
definition can be expressed, whatever it may be.

There are no other known examples of life beyond the Earth. Further, all known
extant and extinct forms of life are generally thought to share a universal
common ancestory (UCA), an idea which was first suggested by
\citet{darwin:1859} and today forms a central pillar of modern evolutionary
theory \citep{sober:2008}. We are therefore faced with the challenge of trying
to infer the rate of abiogenesis from a single datum. Further, we cannot
exclude the possibility than multiple abiogenesis events occurred or are indeed
in the process of occuring, merely that at least one event transpired which led
to a planet-dominating tree of life. A statistic of one is difficult to work
with at the best of times, but the situation is excerbated by the anthropic bias
that we would not be here had at least one event not been successful.

A critical piece of observational evidence guiding these discussions is that
life appears to have emerged on the Earth over 3.5 billion years ago
\citep{schopf:2007}, and thus relatively early. From this observation, there
has been numerous efforts to quantify $\lambda$ conditioned upon this datum,
with, for example, \citet{carter:1983} concluding it is compatible with life
being rare, whereas \citet{lineweaver:2002} finding it points strongly towards
life being common on Earth-like worlds.

A significant advancement in the statistical treatment was presented in
\citet{spiegel:2012}, who cast the problem in a Bayesian framework. As we show
later in a re-derivation of their formalism, this naturally accounts for the
selection bias imposed by the fact our existence is conditioned upon a success.
A key conclusion of \citet{spiegel:2012} was that the probability distribution
of $\lambda$ conditioned upon the observation of life's early emergence, known
as the posterior, is highly sensitive to what prior assumptions (including
tacit assumptions) one imposes about how $\lambda$ is distributed, known as the
prior.

In general, the functional form and corresponding shape parameters describing
the prior probability distribution for $\lambda$ are completely unknown. Even
changing between different but ostensibly reasonable uninformative priors
(e.g. see \citealt{lacki:2016}) leads to dramatic changes in the posterior
\citep{spiegel:2012}, given the current data in hand.

Rather than try to estimate the absolute posterior of $\lambda$, which may
be simply infeasible with present information, a more achievable goal is to
compare how the posterior would be expected to change if additional information
were obtained - a distinct approach. By quantifying how much we learn from
various hypothetical experiments, the goal here is to guide future
experimenters as to which strategies are most likely to constrain $\lambda$,
and critically \textit{in what way} does each experiment constrain it.
Accordingly, in this work, we consider three types of experiments that might be
expected to constrain $\lambda$:

\begin{itemize}
\item[{\tiny$\blacksquare$}] An earlier estimate for the time at which life
first appeared on the Earth, $t_{\mathrm{obs}}$.
\item[{\tiny$\blacksquare$}] An upper limit on the rate of abiogenesis,
$\lambda_{\mathrm{max}}$ (e.g. from a series of \citet{miller:1959} type
laboratory experiments).
\item[{\tiny$\blacksquare$}] The unambigious detection or null-detection of life
amongst a subset of $N$ Earth-like exoplanets/exomoons.
\end{itemize}

Observational evidence for an earlier $t_{\mathrm{obs}}$ may come from numerous
different sources but the details are unimportant for this work - we only assume
that the evidence is unambiguous. This means that we do not invoke a
probabilistic model where the data itself are uncertain, since the probability
distribution describing that uncertainty would be itself unknown.

In a similar vein, an upper limit on $\lambda$ from laboratory experiments is
not a soft probabilistic threshold but assumed to be a hard cut off, since
again the softness shape would add additional unknowns into our model. Further,
we do not consider the more informative case of an actual instance of
spontaneous lab-based abiogenesis in this work. If reproducible, such a
detection would be so constraining it would make the results of our paper moot,
and is therefore not worth including here.

Finally, extraterrestrial examples of life are formally (but not necessarily
strictly) limited to extrasolar worlds (i.e. not endoworlds such as Mars),
in order to greatly reduce the false-alarm probability that the observed life
is in fact related to us via a prior panspermia event. We also assume that that
it is possible to collect observations that can categorically determine whether
an extrasolar body has or does not have recognizeable life. In this way, one
might limit our definition of $\lambda$ to being only applicable to events
which ultimately lead to recognizeable life. For example, it is possible that a
body is inhabited by a life form which produces almost no biosignatures making
remote detection effectively impossible and our $\lambda$ rate would not
include these more elusive organisms. A final caveat is that the planets/moons
in this sample should be not only Earth-like now but Earth-like for geological
timescales (we formally adopt 5\,Gyr as a representative habitable age).

In each of these three experiments, we can compare how the posterior
distribution for $\lambda$ is expected to change, both qualitatively and
quantitatively. Before doing so, it is necessary to first formally introduce
our Bayesian framework in Section~\ref{sec:model}. Following this, we
describe the results of the three experiments in Section~\ref{sec:results}.
Finally, Section~\ref{sec:discussion} summarizes our findings and describes the
main conclusions of this work, as well listing the limitations of this
excercise.

\section{Bayesian Model}
\label{sec:model}

\subsection{A uniform rate model for abiogenesis}
\label{sub:poisson}

We follow the prescription of \citet{spiegel:2012} in adopting the simple yet
reasonable assumption that abiogenesis events occur at a uniform rate over
time, thus following a Poisson process. Let us denote the Poisson shape 
parameter as $\lambda$, which describes the mean number of abiogenesis events,
$N$, ocurring in a fixed time interval, which we set to a Gyr. For example, a
rate of $\lambda=4$ means that, on average, four abiogenesis events
occur every Gyr. The probability distribution for the number of events that
actually occur is described by a Poisson distribution, with a shape parameter
$\lambda$; i.e. $N \sim \mathrm{Poisson}(\lambda)$. Over a time interval of
$t$ then, the expectation value for the number of abiogenesis events 
would be $\lambda t$. Consequently, the probability of $N$ abiogenesis events
having transpired during a time interval $t$ is

\begin{align}
\mathrm{Pr}(N | \lambda,t) &= e^{-\lambda t} \frac{(\lambda t)^N}{N!}.
\end{align}

\subsection{Time for life to emerge as an exponential distribution}
\label{sub:exponential}

As discussed in Section~\ref{sec:intro}, we do not know how many abiogenesis
events have occured on Earth to date, only that at least one must have occurred
over a time interval $t$, where $t$ represents the interval from now back to
when the Earth first became capable of
supporting life. The probability of at least one abiogenesis event occuring in
this time is unity minus the probability of zero events occuring; or
equivalently the probability of life emerging on a planet within a time $t$ is

\begin{align}
\mathrm{Pr}(N\geq1 | \lambda,t) &= 1 - \mathrm{Pr}(N=0 | \lambda,t), \nonumber\\
\qquad&= 1 - e^{-\lambda t},
\label{eqn:cdf}
\end{align}
	
which is the cumulative density function (CDF) of an exponential distribution.
When we consider that this CDF describes the probability that life arose by a
time $t$, it therefore follows that the probability distribution for the time
at which life first arose, which we can denote as $\tlife$, must be an
exponential distribution i.e. $t_{\mathrm{life}}\sim\mathcal{E}(\lambda)$.

\subsection{Accounting for selection bias}
\label{sub:truncation}

For the moment, let us treat $t_{\mathrm{life}}$ as a random
variable drawn from $\mathcal{E}(\lambda)$ with no other constraints
whatsoever on its value (i.e. we will not yet include the paleotonological
information regarding the earliest evidence for life).
Consider setting $\lambda$ to be some low rate, such as $0.01$\,Gyr$^{-1}$.
In such a case, the vast majority of random draws from the the probability
distribution describing $t_{\mathrm{life}}$ will exceed $\sim4.5$\,Gyr.
Even if the Earth can be considered to be habitable immediately after the
hypothesized Moon-forming impact 4.47\,Gyr ago \citep{bottke:2015}, clearly
draws exceeding this age are incompatible with humanity's existence,
else life should not have arisen yet. Accordingly, one might justifiably
include this constraint into our statistical treatment by setting the
maximum limit on $t_{\mathrm{life}}$ to be less than $\sim4.5$\,Gyr.

More generally, one might reasonably argue that 4.5\,Gyr is too optimistic,
and that following the Moon-forming impact, it would have taken some time
for conditions to be suitable for life - for example a crust likely did
not condense until 4.4\,Gyr ago \citep{valley:2014}. On the other hand, one
might argue that even if conditions were immediately suitable for life,
4.5\,Gyr is still too generous a maximum limit, since had life begun just
0.5\,Gyr ago, there would have been insufficient time for intelligent
observers, such as ourselves, to have evolved. We may absorb these arguments
into a single term, $\tau$, which describes the latest time for which life
could have arisen on the Earth and yet still be compatible with both the
habitable history of our planet, and the time it would take intelligent
observers such as ourselves to evolve. The selection bias introduced by $\tau$
can be formally encoded into our model by applying a truncation to the PDF for
$\tlife$, such that

\begin{align}
\mathrm{Pr}(\tlife;\tau) &=
\frac{ \lambda e^{-\lambda \tlife} }{ 1 - e^{-\lambda \tau} }.
\label{eqn:truncpdf}
\end{align}

\subsection{Accounting for observation time}
\label{sub:tobs}

Consider that we have some measurement for the existence of the earliest life
on Earth which occurs at a time $t=\tobs$. We set $t=0$ to be the time at
which the planet became habitable and thus $\tobs$ is perhaps better thought of
as the difference between these two times. We set the reference time this way
in order to remove a degree of freedom from our model, in contrast to
\citet{spiegel:2012} who leaves this in as an extra unknown. With our reference
time, $\tau$ is directly interpretted as the minimum time it takes for life
to evolve from whatever biological entity emerged from the abiogenesis event to
an intelligent observer.

The measurement of $\tobs$ is our most constraining datum but the emergence of
life must in fact predate this time, such that $\tlife\leq\tobs$. This
constraint can be encoded in our probability framework by evaluating the
probability of life emerging before a time $\tobs$:

\begin{align}
\mathrm{Pr}(\tlife\leq\tobs;\tau) &= \int_{\tlife=0}^{\tlife=\tobs} \frac{ \lambda e^{-\lambda \tlife} }{ 1 - e^{-\lambda \tau} } \,\mathrm{d}\tlife,
\end{align}

which, after simplification, gives

\begin{align}
\mathrm{Pr}(\tlife\leq\tobs;\tau) &= \frac{1-e^{-\lambda \tobs}}{1-e^{-\lambda \tau}}.
\label{eqn:trunccdf}
\end{align}

\subsection{Likelihood function for $\lambda$}
\label{sub:likelihood}

Consider that we wanted to infer $\lambda$, conditioned upon a measurement of
$\tobs$, in words we wish to infer

\begin{align}
\underbrace{\mathrm{Pr}(\lambda|\tobs)}_{\mathrm{posterior}} &=
\frac{ \overbrace{\mathrm{Pr}(\tobs|\lambda)}^{\mathrm{likelihood}} \overbrace{\mathrm{Pr}(\lambda)}^{\mathrm{prior}} }{ \underbrace{\mathrm{Pr}(\tobs)}_{\mathrm{evidence}} },
\end{align}

The likelihood function is given by Equation~\ref{eqn:trunccdf}, or re-writing in
a conditional form:

\begin{align}
\mathrm{Pr}(\tobs|\lambda) &= \frac{1-e^{-\lambda \tobs}}{1-e^{-\lambda \tau}},
\end{align}

which is the same result obtained by \citet{spiegel:2012}. This is not
surprising given we began from the same basic assumption of a uniform rate
model for abiogenesis, but we hope our independent take on the derivation
more clearly explains it's origins. It is worth noting that the likelihood
function has a maximum at $\lambda\to\infty$ and displays asymptotic
behavior towards that limit.

\subsection{Multiplanet likelihood function for $\lambda$}
\label{sub:multilike}

Thus far, we have presented a Bayesian model for inferring the $\lambda$
conditioned upon evidence for life on the Earth by a time $\tobs$. An extension
to the above, which was not derived in \citet{spiegel:2012}, is to consider
surveying $N$ Earth-like worlds for life and detecting $M$ examples of it.

If we surveyed planets and moons in the Solar System, there is a plausible
causal connection between the bodies via panspermia \citep{wessen:2010}, which
would significantly complicate the analysis. Instead, we focus on exoplanets
where it can be reasonably assumed that each exoplanetary \textit{system} is an
independent sample. Generally, in what follows, we refer to each exoplanet
system as simply a planet for brevity, but this can include systems of
planets and moons. The only conditional is that only worlds surveyed belong to
a subset of worlds which share similar properties to the Earth i.e. Earth-like.

One difference to our earlier framework is that each exoplanet will typically
have a fairly weak age constraint and an even worse constraint on $\tobs$. This
simplifies our analysis though, since we can approximately assume that all
planets in the sample share the same age and life simply arose before that age
at any time. 
We set this fiducial age to be 5 Gyr, comparable to the age of the Solar system.
We make the further assumption that
$\lambda$ is a common value to all planets surveyed. As discussed in
Section~\ref{sec:intro}, an assumption of this type is fundamentally necessary
to make further progress, but can be considered to be justifiable when the survey
sample belong to a class of Earth-like worlds.

The probability that any one of these planets is observed to be a positive
detection is given by Equation 2, $p = 1-e^{-\lambda t}$, i.e. the likelihood function
defined for the single planet case. 
For simplicity, we assume that all $N$ exoplanets share the same probability
of positive detection in each experiment.
For each set of N exoplanets, we draw a $\lambda$ from its prior and perform
$N$ Bernoulli experiments with success probability $p$.
The probability of obtaining $M$ successes from
a total of $N$ Bernouilli experiments describes a Binomial distribution and
thus we may write that

\begin{align}
\mathrm{Pr}(N,p;M) &= (1-p)^{N-M} p^M \binom{N}{M}.
\end{align}

In the limit of $N=1$ and $M=1$, we get back the same result for the
single-planet case, as expected:

\begin{equation}
\lim_{N\to1} \mathrm{Pr}(N,p;M) =
\begin{cases}
p		& \text{if } M = 1 ,\\
1-p		& \text{if } M = 0 .
\end{cases}
\end{equation}

\subsection{Choosing priors for $\lambda$ and $\tau$}
\label{sub:priors}

A key result of \citet{spiegel:2012} is that changing the prior on $\lambda$
has a major effect on its posterior. As discussed in Section~\ref{sec:intro},
this makes the goal of an absolute determination of the posterior somewhat
unachievable, unless we knew what the correct prior was. However, this work
is chiefly concerned with how the posterior changes when new information is
acquired, seeking the relative differences between hypothetical posteriors
rather than the absolute truth. For this reason, the choice of prior is less
crucial than before and we elect to adopt a simple objective prior in the
form of a log uniform distribution because it is always positive, admits a wide 
range of order of magnitudes, and is non-informative with the maximum entropy (in log space):

\begin{equation}
\mathrm{Pr}(\lambda) = \frac{1}{\lambda} \frac{1}{\log \lambda_{\mathrm{max}} - \log \lambda_{\mathrm{min}}}.
\label{eqn:priorlambda}
\end{equation}

We highlight that this was one of the candidate priors considered by
\citet{spiegel:2012} too. For $\lambda_{\mathrm{min}}$, we fix it at
$10^{-3}$\,Gyr$^{-1}$ in the majority of simulations that follow. In
contrast, \citet{spiegel:2012} considered three different values of
$10^{-3}$\,Gyr$^{-1}$, $10^{-11}$\,Gyr$^{-1}$ and $10^{-22}$\,Gyr$^{-1}$,
which roughly correspond to life occuring once once per 200 star, once in
our galaxy and once in the observable Universe (assuming one Earth-like
planet per star). We will return to these other choices in later in
Section~\ref{sec:results}. The effect $\lambda_{\mathrm{max}}$ will be
investigated in detail as one of our three hypothetical experiments
and so we also discuss this later.

The final term we require a prior for is $\tau$. The largest allowed value
for this term is well-defined as being the age of the Earth, 4.5\,Gyr. The
most conservative estimate for the first life on Earth is $\sim$3\,Gyr ago as \citet{simpson:2016}
suggested a 3\,Gyr minimum time for life to evolve from the very basic form
to being intelligent as we are, which leads to a lower limit of tau to be 1.5\,Gyr. 
Thus the least informative prior is a uniform distribution
given by

\begin{equation}
\mathrm{Pr}(\tau) = \mathrm{U}[1.5,4.5],
\end{equation}

and this is the distribution adopted in what follows.

\subsection{Sampling method}
\label{sub:sampling}

From a sampling perspective, our model has relatively compact dimensionality
and has only a single data point. In such a case, we are able to employ a
highly efficient sampling algorithm known as the ``bootstrap filter'', which
is a class of ``particle filter'' \citep{kunsch:2013}.

To briefly describe the algorithm, first $N$ sets of hyper-parameters
($\Theta_H^1, \Theta_H^2,..., \Theta_H^N$) are drawn according to the hyper
prior distribution. $N$ sets of local parameters are then also
drawn given the hyper-parameters ($\Theta_L^i,\ i=1,..,N$). Next, we calculate
the likelihood of our data given the parameters ($L^i,\ i=1,..,N$). Finally, we
resample parameters $(\Theta_H,\Theta_L)$ with their corresponding likelihoods
$L$ as weights. 

In order for the final sampled parameters to be well spread over the parameter
space, the parameter space need to be well explored in the first step. This
indicates that the method would not work for high dimensional problems,
since the required number of samples would be intractable in such a parameter
space. In such cases, other sampling methods, like Markov Chain Monte Carlo
\citep{metropolis:1953} would be more suitable. But clearly for a simple model
like the one we have, the bootstrap filter sampling method is well-suited and
highly efficient. 

\subsection{Using information gain to compare different posteriors}
\label{sub:KLD}

The goal of the paper is to compare how the $\lambda$ posterior changes with
different experimental setups. To evaluate the difference in a quantitative
way, we choose to use the Kullback-Leibler divergence (KLD), which is also
known as relative entropy \citep{KLD:1951}. It quantifies the entropy change
in going from probability distribution $Q$ to $P$, where a result of zero
implies no difference and non-zero (but always positive) values imply a
finite difference, and is given by

\begin{align}
\mathrm{KLD}(\mathcal{P}||Q) &= \int \mathcal{P}(x) \log \frac{\mathcal{P}(x)}{Q(x)}\,\mathrm{d}x,
\label{eqn:KLD}
\end{align}

where the integrand limits cover the complete supported domain of the
functions $P$ and $Q$. We highlight that the popular Kolmogorov–Smirnov
\citep{kolmogorov:1933,smirnov:1948} or Anderson-Darling \citep{AD:1952}
tests are primarily designed to test for when two distributions show
significant departures from each other and thus would not be directly
applicable here. A test statistic, such as the Kolmogorov-Smirnov distance,
could be utilized to quantify differences although we note that this metric
only codifies the maximum difference in the CDF between two distributions,
and does not fully account for the ensemble of differences occuring across
the parameter space. For these reasons, we ultimately concluded that the KLD
would be a well-suited tool for quantifying the differences observed in our
hypothetical experiments.

Computationally we need to use the discretized version of the equation to
calculate the information gain. For this work, we use an R package
{\tt entropy} with all the KLD calculations. However, as the results of the
paper shows, we should not only depend on a number to tell the difference
between two distributions. It is useful within each experiment. But when we
update the posterior with different experiments, thus changing the posterior
in different ways, a number is not enough to describe the results.

\section{Results}
\label{sec:results}
 
\subsection{Experiment 1: Reducing $t_{\mathrm{obs}}$}
\label{sub:paleo}

\subsubsection{Overview}

With our model and objective established, we now describe the results from our
hypothetical experiments, starting with reducing $t_{\mathrm{obs}}$. We
therefore consider here, what effect would an earlier estimate for the first
life on Earth have on our knowledge concerning the rate of abiogenesis events,
$\lambda$?

The earliest undisputed evidence for life on Earth comes from $\sim$3465\,Myr
Archean deposits in the Apex Basalt of Western Australia, containing morphotype
units which are concluded to be microfossils in \citep{schopf:2007}. Given that
the Earth formed $(4.54\pm0.05)$\,Gyr ago \citep{dalrymple:2001}, then the
maximum plausible value we can assign to $t_{\mathrm{obs}}$ would be
$\simeq$1\,Gyr.

In contrast, the very earliest claim for the first evidence of life extends
as far back as 4280\,Myr ago \citep{dodd:2017}, from putative
fossilized microorganisms in ferruginous sedimentary rocks from the
Nuvvuagittuq belt in Quebec, Canada. The study of ancient zircons in Jack
Hills, Western Australia, indicates the presence of oceans on the Earth as far
back as $(4408\pm8)$\,Myr \citep{wilde:2001}, and so one might argue from
these studies that $t_{\mathrm{obs}}$ could be as short as $\sim100$\,Myr.

As evident from the cited literature, this is an active and rapidly developing
field and thus it is quite likely that further revisions to $t_{\mathrm{obs}}$
may occur in the near future. If the age is revised down by another factor
of ten, though, how much does this really teach us about $\lambda$? The intuitive
temptation is to assign earlier start dates with evidence that life is not
fussy and can start quite easily i.e. a high $\lambda$ \citep{allwood:2016}.
As discussed earlier, this question can be more readily addressed in a Bayesian
informatics framework such as that presented here.

To investigate this, we computed the posterior for several thousand different
hypothetical values of $t_{\mathrm{obs}}$ log-uniformly spaced between
$10^0$\,Gyr (corresponding to the modern conservative limit) down to
$10^{-3}$\,Gyr (a deliberately highly optimistic choice). In each simulation,
we fix $\lambda_{\mathrm{min}} = 10^{-3}$\,Gyr$^{-1}$, as discussed in
Section~\ref{sub:priors}, but explore four different candidate values for
$\lambda_{\mathrm{max}}$ of $10^{0}$\,Gyr$^{-1}$, $10^{+1}$\,Gyr$^{-1}$,
$10^{+2}$\,Gyr$^{-1}$ and $10^{+3}$\,Gyr$^{-1}$.

\subsubsection{Qualitative impacts on the $\lambda$ posterior}

\begin{figure*}
\begin{center}
\includegraphics[width = 15.0cm]{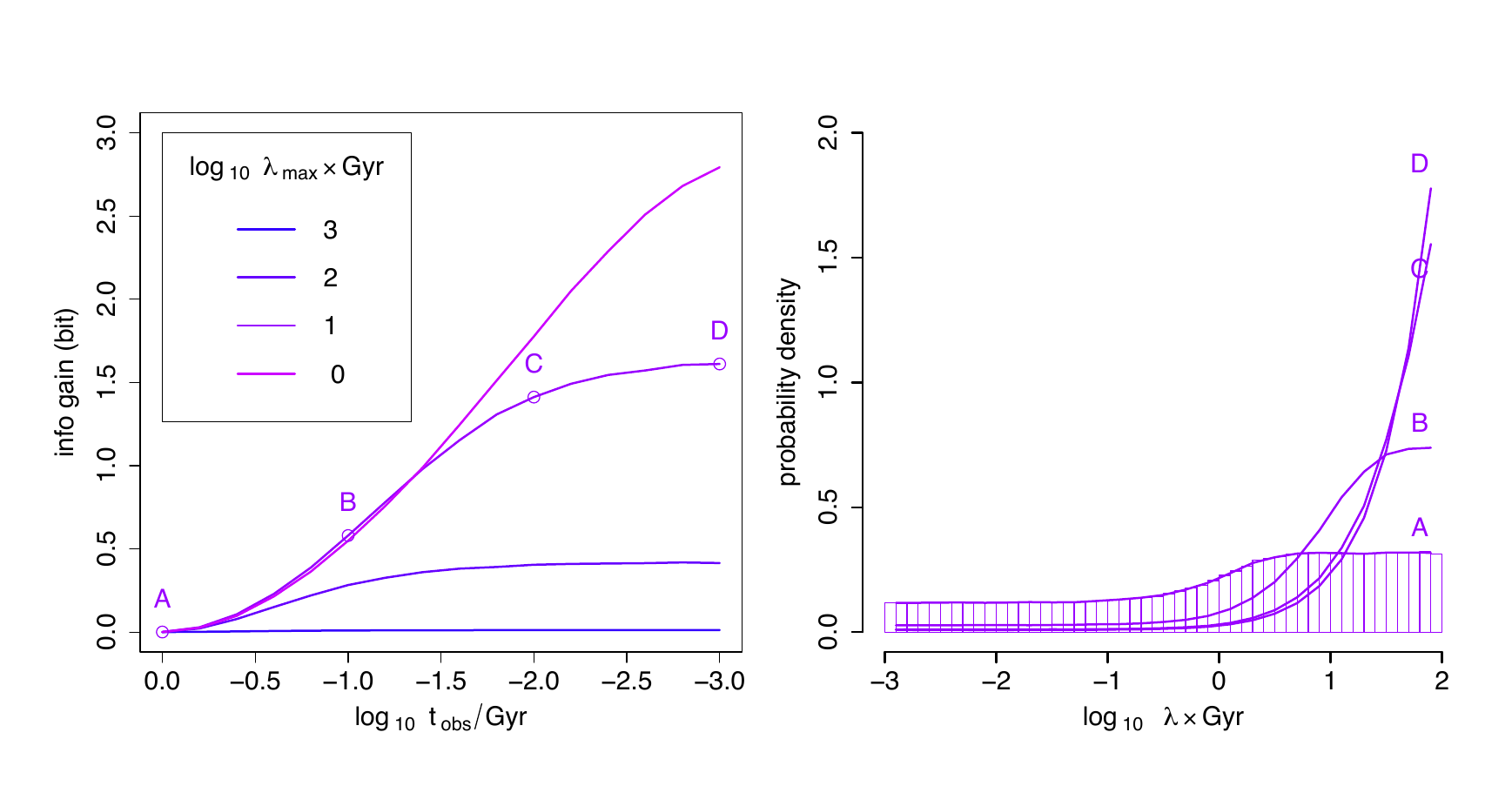}
\caption{The left panel shows information gain calculated by KLD as
$t_{\mathrm{obs}}$ gets smaller. The right panel gives the detailed posterior
distribution of $\lambda$ for fixed $\lambda_{\mathrm{max}}$ of
$10^2$\,Gyr$^{-1}$. Line A, B, C, D represents $t_{\mathrm{obs}}$ of
$10^0$\,Gyr, $10^{-1}$\,Gyr, $10^{-2}$\,Gyr and $10^{-3}$\,Gyr respectively.}
\label{fig:paleo}
\end{center}
\end{figure*}

In the right panel of Figure~\ref{fig:paleo}, we present the posterior
distributions of $\lambda$ for all four particular $t_{\mathrm{obs}}$
choices when $\lambda_{\mathrm{max}}=10^{+2}$\,Gyr$^{-1}$, to illustrate
the qualative differences which arise. Broadly speaking, it can be seen
that as $t_{\mathrm{obs}}$ becomes smaller, a higher value of the abiogenesis
rate is favored. In all cases, the posterior is a monotonic function
consistent with a lower limit constraint on $\lambda$ rather than a peaked,
quasi-Gaussian measurement. This immediately reveals that measurements
of $t_{\mathrm{obs}}$ alone can only hope to ever place probabilistic lower
limits on $\lambda$, and never produce what might be considered as a direct
measurement.

When $t_{\mathrm{obs}}=10^0$\,Gyr (line A), which corresponds to the
conservative limit, the posterior stays almost flat after a rise at around
$\lambda=10^0$\,Gyr$^{-1}$ (i.e. one event per Gyr). This can be
understood by the observation that we can't distinguish between, for example,
10 or 100 events per Gyr, as they are both sufficient to have
$t_{\mathrm{obs}}=10^0$\,Gyr. Accordingly, the posterior largely follows the
prior. More generally, one expects a inflection point in the posterior to occur
at $\lambda \simeq \lambda_{\mathrm{inflection}} = 
t_{\mathrm{obs}}^{-1}$\,Gyr$^{-1}$.

As $t_{\mathrm{obs}}$ gets larger, higher rate becomes clearly more favorable.
The observed posterior morphpoligies are also responsive to the fact that we
set $\lambda_{\mathrm{max}}$ at $10^2$\,Gyr$^{-1}$. If say, it is set were as
high as $10^5$\,Gyr$^{-1}$, we should expect all four lines to reach a plateau
(which we indeed verified). Accordingly, one can conclude that the actual
posterior distribution is highly sensitive to whatever value one adopts for the
prior's upper limit, $\lambda_{\mathrm{max}}$, re-inforcing the conclusions of
\citealt{spiegel:2012}.

\subsubsection{Quantitative impacts on the $\lambda$ posterior}

The left panel of Figure~\ref{fig:paleo} show the information gain obtained
from revising $t_{\mathrm{obs}}$ down. Specifically, we set $Q$ in
Equation~\ref{eqn:KLD} to be the posterior distribution of $\lambda$ when
$t_{\mathrm{obs}}  = 10^0$\,Gyr, corresponding to our current observational
constraint. We do not use the prior on $\lambda$, since our ``starting point''
in terms of current knowledge includes a measurement of $t_{\mathrm{obs}}$,
and that should be incorporated in our comparison. Accordingly, $P$ in
Equation~\ref{eqn:KLD} is now set to be equal to the derived posteriors at
other choices of $t_{\mathrm{obs}}$. As a result of this setup, the information
gain between $P$ and $Q$ when $t_{\mathrm{obs}}=1$\,Gyr is, by definition,
zero.

The four lines shown broadly reproduce expectation. The information gain
increases, resembling a non-linear logistic function, as $t_{\mathrm{obs}}$ is
revised down. Clearly the choice of $\lambda_{\mathrm{max}}$ again has a
significant impact and specifically seems to control a saturation limit in the
information gain. This is most clearly seen for the curve describing
$\lambda_{\mathrm{max}}=10^1$\,Gyr$^{-1}$, where there is negligible
information gain once $t_{\mathrm{obs}}$ drops below $10^{-1.5}$\,Gyr. Indeed,
if $\lambda_{\mathrm{max}}=10^0$\,Gyr$^{-1}$, even our current conservative
limit on $t_{\mathrm{obs}}$ lives on this saturated plateau. In such a
circumstance, there is essentially no value in paleontologists continueing to
try and revise $t_{\mathrm{obs}}$ further back (if their goal is to constrain
$\lambda$).

The reason behind the satuaration can be explained by careful examination of
the right panel plot. In the case of line A, the first abiogenesis event
happened at $t_{\mathrm{obs}}=1$\,Gyr. Since there is an inflection point in
the posterior at $\lambda_{\mathrm{inflection}} = 
t_{\mathrm{obs}}^{-1}$\,Gyr$^{-1}$ (as established earlier), then the posterior
experiences a steep rise at around once per Gyr. Likewise, line B/C/D should
rise at around $10^1$\,Gyr$^{-1}$, $10^2$\,Gyr$^{-1}$ and $10^3$\,Gyr$^{-1}$
respectively. However, for the case of line D, since the upper limit is set at $\lambda_{\mathrm{max}} = 10^2$\,Gyr$^{-1}$, much of its morphological changes
are truncated. This upper limit constraint forces line D to perform similarly
like line C. If we translate that to information gain, we can see that point D
doesn't increase very much from point C, and the whole line will gradually
saturate.  

\subsection{Experiment 2: Reducing $\lambda_{\mathrm{max}}$}
\label{sub:millerurey}

\subsubsection{Overview}

The second type of experimental pressure we consider on $\lambda$ is one driven
by lab-based experiments. In particular, we consider the thought experiment
where a large suite of containers are constructed, within each exists a
representative environment of an Earth-like planet and simply count how often
does life spontaneously emerge from these containers. This is not meant to be
a practical experimental setup but rather a toy example of how one might
construct a series of experiments to constrain $\lambda$ in the lab.

In principle, one or more of these experiments might successfully spawn a new
form of life. If this result were reproducible and verifiable, it would provide
such dramatic insights into abiogenesis that comparing it to the other
hypothetical experiments in this work is somewhat of a moot point. Instead, we
take the pessimistic angle that the experiments do not successfully produce
a single abiogenesis event (which is of course consistent with current
experiments). In such a case, one could reasonably infer an upper limit on
$\lambda$ from the null results and thus we envisage that these experiments
return a single new datum for our setup - $\lambda_{\mathrm{max}}$.

It is therefore instructive to compare how obtaining ever tighter limits on
$\lambda_{\mathrm{max}}$ affects the posterior on $\lambda$ without changing
$t_{\mathrm{obs}}$. To do so, we varied $\lambda_{\mathrm{max}}$ log-uniformly
across the same range as used in Section~\ref{sub:paleo}, from
$10^{3}$\,Gyr$^{-1}$ to $10^0$\,Gyr$^{-1}$ but now at a much finer resolution.
In each simulation we derive the resulting $\lambda$ posterior and repeat the
entire excercise for four different choices of $t_{\mathrm{obs}}$, namely
$10^0$\,Gyr, $10^{-1}$\,Gyr, $10^{-2}$\,Gyr and $10^{-3}$\,Gyr. Thus, our
parameter range directly mirrors the range considered in
Section~\ref{sub:paleo}. As before, $\lambda_{\mathrm{min}}$ is fixed to $10^{-3}$\,Gyr$^{-1}$.

\subsubsection{Qualitative impacts on the $\lambda$ posterior}

\begin{figure*}
\begin{center}
\includegraphics[width = 15.0cm]{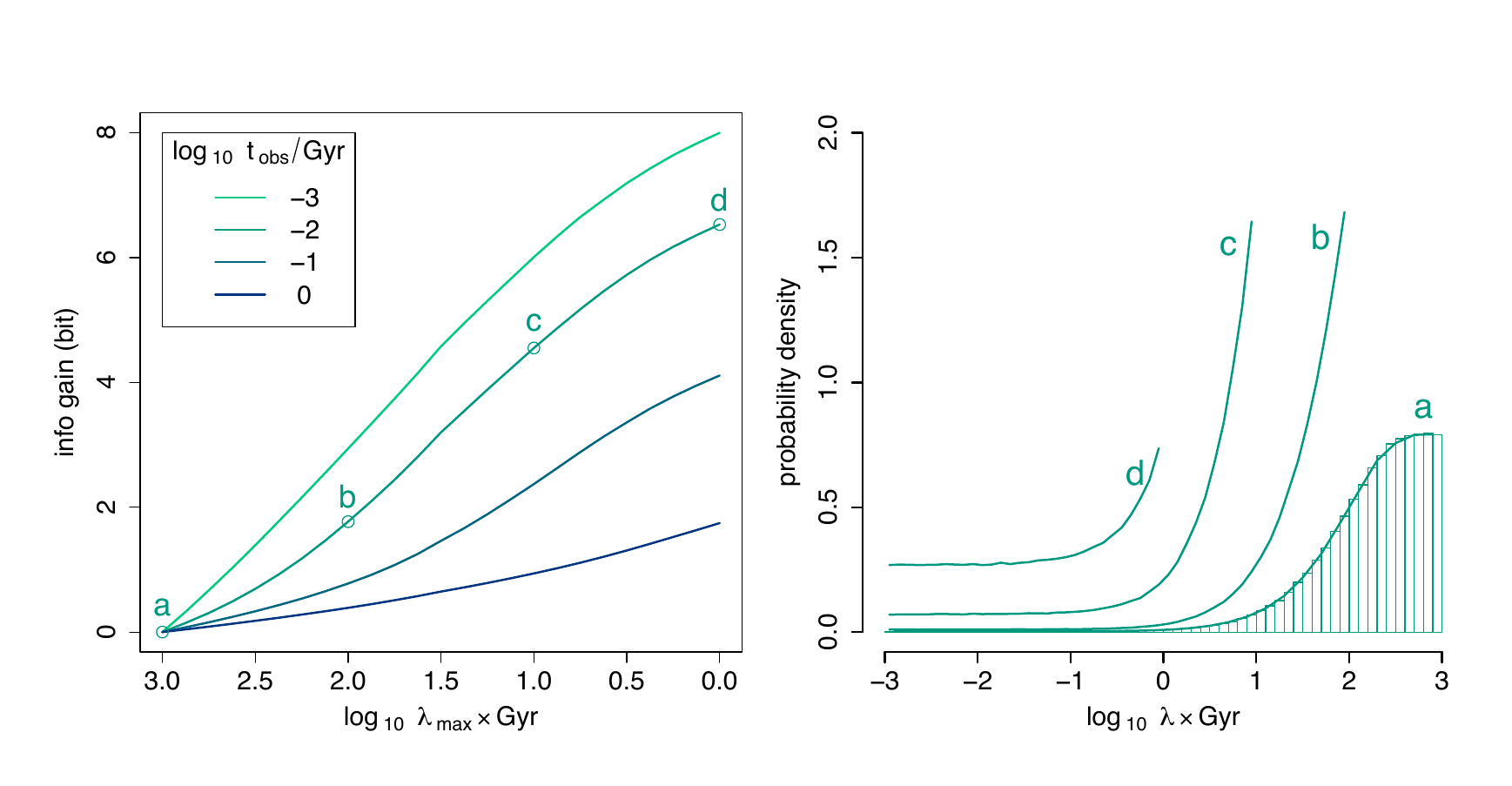}
\caption{The left panel shows information gain calculated by KLD as
$\lambda_{\mathrm{max}}$ gets smaller. The right panel gives the detailed
posterior distribution of $\lambda$ for fixed $t_{\mathrm{obs}}$ of
$10^{-1}$\,Gyr. Line a, b, c, d represents $\lambda_{\mathrm{max}}$ of
$10^3$\,Gyr$^{-1}$, $10^2$\,Gyr$^{-1}$, $10^1$\,Gyr$^{-1}$ and
$10^0$\,Gyr$^{-1}$ respectively.}
\label{fig:millerurey}
\end{center}
\end{figure*}

The right panel of Figure~\ref{fig:millerurey}, we show four examples of how
the $\lambda$ posterior changes for different $\lambda_{\mathrm{max}}$ values,
keeping a fixed $t_{\mathrm{obs}}=10^{-2}$\,Gyr. It is clear that the main
difference between the updated posteriors (line b, c, d) and the original
posterior (histogram/line a) is that the posterior becomes truncated at
smaller $\lambda$ values, corresponding directly to $\lambda_{\mathrm{max}}$.
As a result, whilst line a shows a plateau, this is eroded by the truncation
of the sharper $\lambda_{\mathrm{max}}$ constraints, for similar reasoning as
that discussed in Section~\ref{sub:paleo}.

\subsubsection{Quantitative impacts on the $\lambda$ posterior}

In left-panel of Figure~\ref{fig:millerurey} shows the information gain by
varying $\lambda_{\mathrm{max}}$ from $10^3$\,Gyr$^{-1}$ to $10^0$\,Gyr$^{-1}$,
with each of the four lines showing a different assumed $t_{\mathrm{obs}}$
value. In comparison to Figure~\ref{fig:paleo}, the information gains are, in
general, higher in this second experiment. This can be understood to be resulting
from the truncation effect, which significantly elevates the density of lower
$\lambda$ values in order to maintain normalization. Both experiments 1 and 2
can be seen to provide merely lower limits on $\lambda$, rather than a strong
measurement, since they are both ultimately conditioned upon the same datum.
Nevertheless, in a quantative sense, our analysis indicates that there is a 
greater potential for information gain in experiment 2 within the parameter
ranges considered.

This conclusion is re-enforced by the fact that none of our simulations in
experiment 2 led to a neglgible information gain, whereas this was observed in
certain cases for experiment 1. Specifically, once $\lambda_{\mathrm{max}}$
can be constrained to less than or equal to $\sim10^0$-$10^1$\,Gyr$^{-1}$,
there is very little information gain to be had from revising
$t_{\mathrm{obs}}$. In contrast, even the most conservative limit on
$t_{\mathrm{obs}}=10^0$\,Gyr allows for sizable information gains by
revising $\lambda_{\mathrm{max}}$. Further, the remote possibility of an
outright successful abiogenesis event in the lab makes a compelling case
that this enterprise is more likely to teach us about abiogenesis than
experiment 1.

Although we only show lines spanning up to
$\lambda_{\mathrm{max}}=10^3$\,Gyr$^{-1}$, it is worthwhile to consider
the behavior at more extreme choices. As can be seen in
Figure~\ref{fig:millerurey}, line ``a'' saturates but it is only the
line to do so. This is because in this case $\lambda_{\mathrm{max}}$ (which
equals $10^3$\,Gyr) substantially exceeds $\lambda_{\mathrm{inflection}}$
(which equals $10^1$\,Gyr), allowing for the saturation behavior to take
place. We might therefore hypothesize that if we had chosen a far higher
choice of $\lambda_{\mathrm{max}}$, say $10^10$\,Gyr$^{-1}$, then the
posterior would broadly look similar to line ``a'' except the truncation
would occur much later. Accordingly, we might hypothesize that if we
compared the information gain from $\lambda_{\mathrm{max}}=10^10$\,Gyr$^{-1}$
to $\lambda_{\mathrm{max}}=10^9$\,Gyr$^{-1}$, there would be very little
information gain (compared to, say, going from $10^2$\,Gyr$^{-1}$ to
$10^1$\,Gyr$^{-1}$) since in both cases $\lambda_{\mathrm{max}} \gg
\lambda_{\mathrm{inflection}}$. If true, this would mean that if we
extended the left panel plot in Figure~\ref{fig:millerurey} far to the
left (i.e. much higher $\lambda_{\mathrm{max}}$), the curves would be
very flat until we start to encroach upon values in the domain of
$\lambda_{\mathrm{inflection}}$. To verify this hypothesis, we repeated
the experiments up to $\lambda_{\mathrm{max}}=10^10$\,Gyr$^{-1}$ and indeed
verified the information gain curves are approximately flat when 
$\lambda_{\mathrm{max}} \gg \lambda_{\mathrm{inflection}}$.

\subsection{Future Evidence from Exoplanets}
\label{sub:exosurvey}

\subsubsection{Overview}

The third experiment we consider is a future astronomical telescope capable of
discrening unambiguously whether an observed exoplanet hosts life or not.
As with the previous experiments, the specific details of how this is achieved
is not important for the following discussion, although a plausible strategy
would be to seek atmospheric biosignatures \citep{leger:1996}. We note
surveying a large number of Earth-like worlds is beyond the abilities of
existing facilities \citep{seager:2014}, but it is not unreasonable to
suppose that it should be plausible in the future \citep{rauscher:2015}.
If the goal of such an enterprise is to quantify our uniqueness and thus the
rate at which life springs forth on Earth-like worlds, then the thought
experiment described here provides a direct evaluation of how informative such
an effort should be expected to be.

Naturally, prior to having conducted this experiment in reality, the number
of abiogenesis detections, $M$, amongst a sample $N$ exoplanets is unknown,
yet the ratio $M/N$ will clearly strongly affect the resulting $\lambda$
posterior. To account for this, we therefore have two control variables in
this experiment, rather than one: the success rate, $M/N$, and the survey
size, $N$.

\subsubsection{Yield expectations}

It is instructive to first pose the question, what kind of ratio value do
we actually expect, based on current information? A ratio which agrees
with our naive expectation (whatever that may be) would lead to only a small
change in the $\lambda$ posterior and thus a small degree of information
gain. In contrast, a ratio $M/N$ resulting from this hypothetical survey that
in is tension with our prior expectation would dramatically change the $\lambda$
posterior, and thus lead to a large information gain. These considerations
provide some initial insight as to why particular values of $M/N$ may not
necessarily lead to significant gains in our knowledge of $\lambda$. It is
therefore interesting and somewhat counter-intuitive to note that a survey
of $N$ exoplanets for life may not necessarily lead to any substantial gains
in our knowledge about the rate of abiogenesis, depending on what ratio of
success is observed.

From the above arguments, it is clear that an observed $M/N$ close to our
prior expectation on $M/N$ should represent a minimum in the possible
information gain. But what exactly is our prior expectation on $M/N$?
Optimists would say life starts everywhere and thus we expect $M/N\sim1$
\citep{lineweaver:2002}. If so, then detecting a high success rate in
an exoplanet survey would actually teach us very little. Like dropping
a coin and seeing it fall to the ground under gravity, results which
match expectation generally don't teach us as much as if the coin had
travelled upwards. In contrast, a pessimistic might say they are
convinced life is rare and thus any detections elsewhere would be
highly surprising, much like the coin travelling upwards, thereby
teaching us a great deal.

We now turn to estimating what our a-priori expected $M/N$ should be. This is
of course closely related to our current posterior for $\lambda$. However,
as we have argued earlier and indeed as concluded by \citet{spiegel:2012},
an absolute inference of $\lambda$ is not possible unless we know what
the correct prior should be. Although we have investigated the effect of
varying $\lambda_{\mathrm{max}}$ and this may be constrainable via
experiment (see Section~\ref{sub:millerurey}), it is unclear how one
should assign $\lambda_{\mathrm{min}}$ at this time (of course another
choice is the shape of the prior itself, but we leave that aside at this
time).

Using the fiducial choice of $\lambda_{\mathrm{min}} = 10^{-3}$\,Gyr$^{-1}$,
which has been used in both Sections~\ref{sub:paleo} \& \ref{sub:millerurey},
our expectation is that $M/N$ is generally quite high. We demonstrate this
in the left panel of Figure 3, where we compute the average
number of detections expected using our earlier $\lambda$ posteriors on 5\,Gyr
old Earth-like planets (by Monte Carlo experiments drawing Bernoulli trials
using the probability defined in Equation~\ref{eqn:cdf}). For almost any choice
of $t_{\mathrm{obs}}$ or $\lambda_{\mathrm{max}}$ (within the ranges used
theroughout this paper), one can see that $M/N$ is expected to be high.
Accordingly, if we set $\lambda_{\mathrm{min}} = 10^{-3}$\,Gyr$^{-1}$ as
before, experiments where we set $M/N\sim1$ will tend to yield minimal gains in
information content on $\lambda$. We highlight this point carefully due its
somewhat counter-intuitive consequences.

\begin{figure*}
\begin{center}
\includegraphics[width = 15.0cm]{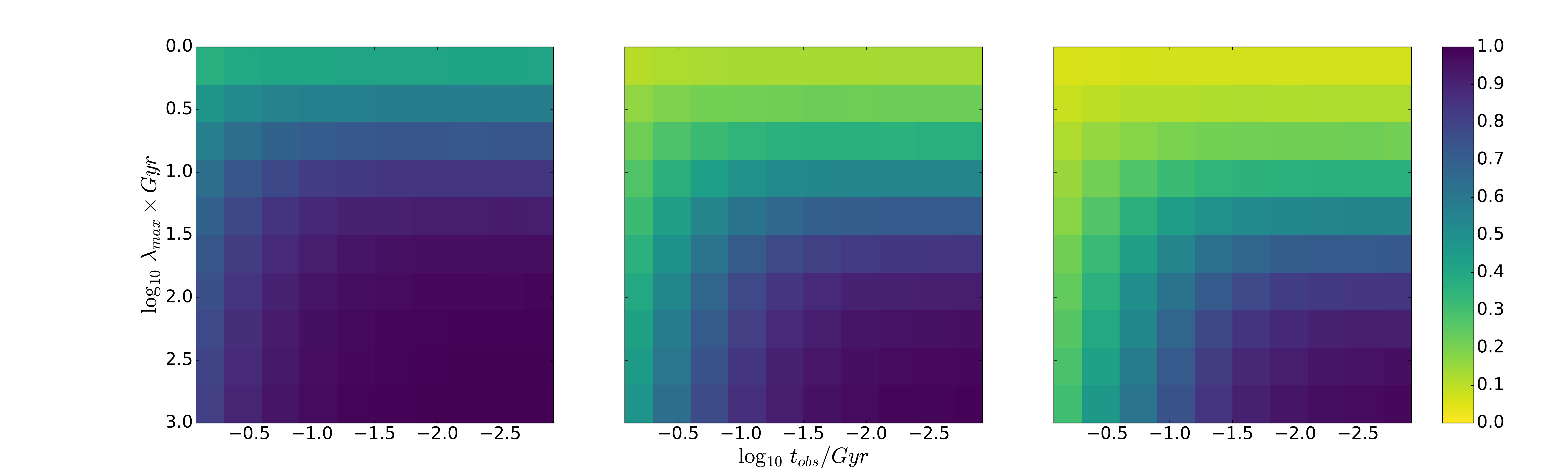}
\caption{The three plots show the predicted ratio of planets with life given
the situation on the Earth. From left to right panel, $\lambda_{\mathrm{min}}
= 10^{-3}$\,Gyr$^{-1}$, $10^{-11}$\,Gyr$^{-1}$, and $10^{-22}$\,Gyr$^{-1}$. In
each panel, we vary $t_{\mathrm{obs}}$ (x-axis) and $\lambda_{\mathrm{max}}$
(y-axis).}
\end{center}
\label{fig:yieldcalc}
\end{figure*}

As Figure 3 makes clear, changing $\lambda_{\mathrm{min}}$
leads to a significant decrease in the expected success yield. For these
reasons, the experiment 3 information gains presented here are
highly sensitive to the assumed values of $\lambda_{\mathrm{min}}$. We note
that this was not the case in experiments 1 and 2, where we verified that
varying $\lambda_{\mathrm{min}}$ to $10^{-11}$\,Gyr$^{-1}$ or
$10^{-22}$\,Gyr$^{-1}$ did not significantly impact the shape of the
posteriors and only slightly affected the scaling of the information gain
plots. For this reason, in what follows, we limit our discussion to being
largely a qualatative one.

\subsubsection{Qualitative impacts on the $\lambda$ posterior and information gain}

\begin{figure*}
\begin{center}
\includegraphics[width = 15.0cm]{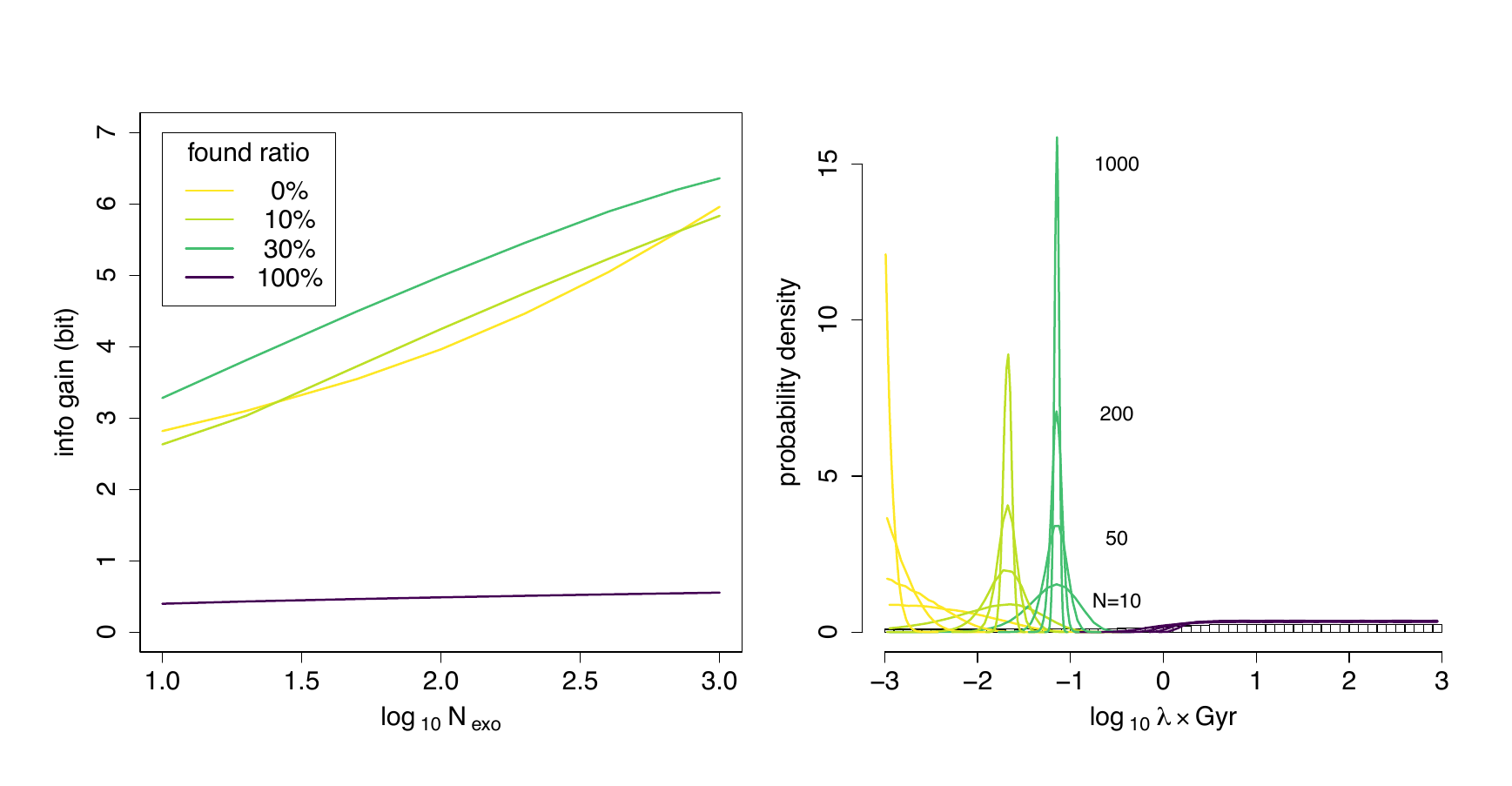}
\caption{The left panel shows information gain calculated as we increase the number of observed exoplanets. The right panel gives the detailed posterior distribution of $\lambda$ for different ratios.}
\label{fig:exosurvey}
\end{center}
\end{figure*}

The right panel of Figure~\ref{fig:exosurvey}, the dark green lines show four
example posteriors for $\lambda$ for $N=10$, $50$, $200$ and $1000$. The dark
green lines all assume $M/N=$30\%. If 30\% of the planets harbor life, then
this implies that a $\lambda$ of $0.3/5$\,Gyr$^{-1}$ and this is indeed where
the posteriors all peak. The peakiness of the posterior naturally increases as
it becomes conditioned upon larger samples of data.

The yellow lines show the case of no detections, $M=0$, which strongly opposes
our prior expectation of a high rate. However, because these are essentially
null detections, the posterior peaks at $\lambda_{\mathrm{min}}$ and does not
converge to a single quasi-Gaussian shape. For this reason, the posterior does
certainly lead to a large information gain but not maximal.

On the other hand, if $M=N$, shown by the dark blue lines, the information gain
is minimal. As explained in detail earlier, this results from our high prior
expectation of $\lambda$ anyway, conditioned upon the Earth's early start to
life. Adding more data only re-enforces this belief leading to minor gains in
information content. Further, as with the $M=0$ case, the posterior peaks at
a prior bound, in this case $\lambda_{\mathrm{max}}$ and thus does not
represent a converged, constrained datum.

The argument made above does not include effects regarding the sample inclusion
though. For example, if all Earth-like planets surveyed show evidence for life,
there would be an opportunity to learn about abiogenesis in a different way
by gently expanding the sample to sub-optimal worlds and observing when the
success rate decreases. This is not formally encoded within our model and
represents just one of the ways that a large sample of exo-life detections
would lead to large gains in understanding of abiogenesis, even if not
significantly improving our inference of the abiogenesis rate of Earth-like
worlds.

\section{Discussion}
\label{sec:discussion}

In this work we have sought to understand how three different experimental
approaches would be expected to inform our knowledge of the rate of
abiogenesis, $\lambda$, on Earth-like worlds. Our approach adopts the Bayesian
formalism of \citet{spiegel:2012} as a means for deriving the probability
distribution of $\lambda$ conditioned upon some observation of the earliest
evidence for life ($t_{\mathrm{obs}}$) and some choice for the prior. As
discussed in \citet{spiegel:2012}, the resulting posteriors are highly
influenced by the choice of prior, and thus an absolute inference of the
posterior is somewhat unachieveable. Instead, our paper focusses on what
the relative gain in information acquired as new experiment evidence is
introduced into the problem across a range of plausible input parameters.

Our Bayesian informatics approach seeks to understand exactly how the $\lambda$
posterior changes in response to new experimental constraints, and whether
there are certain experiments or parameter regimes which can be concluded to
either be negligbely informative, or vice versa greatly informative.

Our work focusses on three thought experiments, which could be plausibly
conducted at present or in the near future: i) paleontological evidence for
an earlier start to life; ii) Monte Carlo experiments seeking to create a
laboratory abiogesis event; and iii) a survey of exoplanets for biosignatures.
To quantify the information gained from each experiment, we employ the
Kullback-Leibler divergence, or relative entropy, to calculate the difference
between the original and the new $\lambda$ posterior. Additionally, we have
performed detailed analyses of the resulting posteriors in an attempt to
understand how their morphologies are sculpted by new constraints.

Without knowledge of the correct limits on the prior (or indeed the shape
of the prior), it is not possible to unambigiously claim that any of these
experiments will always be superior/inferior to the others. Despite this,
there are general trends which emerge from our thought experiments. These
are briefly summarized in Figure 5, and we urge the reader
to explore the more detailed accounts of each found within this paper.

\begin{figure*}
\begin{center}
\includegraphics[width = 15.0cm]{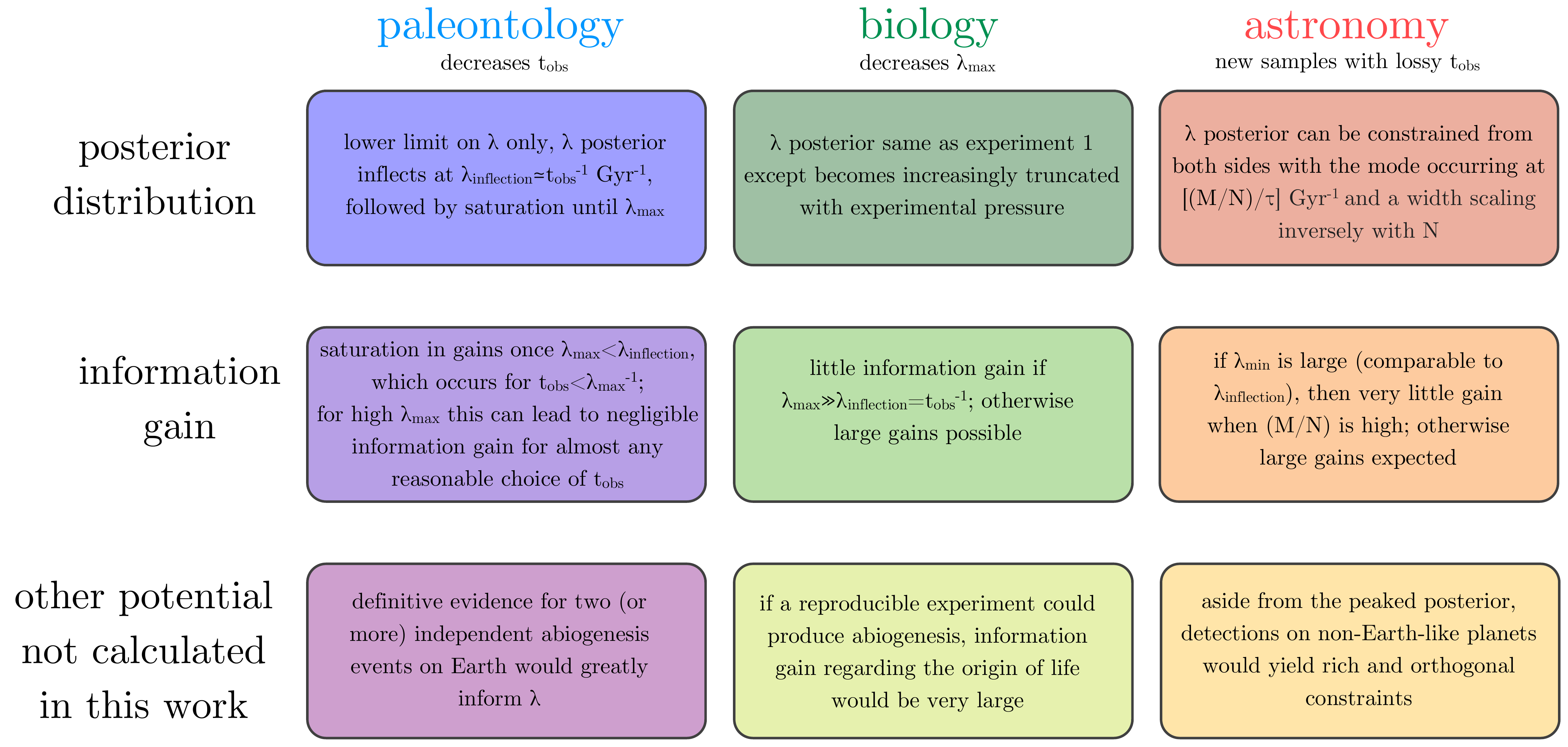}
\caption{Summary chart of some key general results found in this work.}
\end{center}
\label{fig:summary}
\end{figure*}

We highlight a couple of important general trends. First, it is non-intuitive
that an exoplanet survey detecting many instances of life can be highly
uninformative in certain regimes. This is because for certain choices of the
prior, the early start to life on Earth leads to a expectation that all
Earth-like planets are inhabited, thus new detections don't significantly
affect the posterior. As highlighted earlier though, clearly such a result
could be greatly informative if the sample is extended beyond Earth-like
worlds to include more exotic locations, echoing the conclusion of
\citet{lenardic:2018} but for different reasons.

Second, we have ignored the possibility of a successful laboratory abiogenesis
in this work and instead assumed they only yield null-results and thus an
upper limit on $\lambda$. This is motivated by the assumption that the ability
to create life in the lab, in a reproducible experiment, would provide an
enormous amount of information that would make direct comparisons somewhat
meaningless to the other experiments. This is related to the possibility of
paleotology/paleogenetics finding unambigious evidence for a second abiogenesis
at early times, which we also did not explicitly consider.

A general trend we highlight is that unless abiogenesis occurs in the lab,
the bilogical and paleontological experiments are conditioned upon the same
datum - the earliest evidence for life on Earth - and thus both yield only
lower limits on $\lambda$. In contrast, exoplanet surveys will, in general,
yield peaked posteriors constrained from both sides i.e. a measurement
rather than a limit. This is important to emphasize as this is not fully
captured by simply comparing the relative entropies between posteriors.

A final point we emphasize is that paleontology in particular has regimes
in which essentially no information is gained. For reasons discussed in
detail earlier, this occurs when $t_{\mathrm{obs}} \ll 
\lambda_{\mathrm{max}}^{-1}$. Since there is no agreed value for
$\lambda_{\mathrm{max}}$, we are unable to infer what the corresponding
time really is, but it is important to note that even a very early
start to life can, for some choices of $\lambda_{\mathrm{max}}$, make
us none the wiser about abiogenesis.

Being the first effort at a Bayesian informatics analysis of abiogenesis,
we acknowledge that there are outstanding issues that we haven't been able to
address in detail here. First. we did not conduct a full exploration of the
effect of changing $\lambda_{\mathrm{min}}$, largely since we couldn't
conceive of a direct empirical way of constraining it. Another issue is that we
have assumed $\lambda$ follows a universal distribution for all exoplanets and
at all times, while of course we should expect that it will change with different
environments. This was largely done as a simplfying assumption, following on
from \citet{spiegel:2012} and is partly mitigated by our assumption that only
``Earth-like'' worlds are included in an exoplanet survey. This problem might be
better solved in the future with hierarchical models, which were not explored
here. Finally, we have assumed that experiment 3 is actually feasible - which
remains somewhat unclear. Specifically, we assume that life can be unambigiously
detected or ruled out on a planet from remote observations. A suggestion for
future work would be to change this hard binary flag to a softer probability,
implying a more complicated model such as, again a hierarchical model.

Together, we generally find that all of the experiments can certainly
yield constraints on $\lambda$ and in non-overlapping ways. A lab
abiogenesis event would be such an informative experiment that even if it
could be argued that the informations gains from null results are negligible,
there is a strong case to conduct the experiment regardless. Similarly,
early starts to life, whilst they may not formally improve $\lambda$, do
clearly provide information about the conditions which life began and this
is not formally encoded in our model. If an actual measurement of $\lambda$,
is desired, rather than a limit, we would argue that the exoplanet survey
is the most direct way to infer this. Moreover, the ability to expand to
non-Earth like worlds can probe different conditions and thus offer an
orthogonal type of information. Together then, all three experiments deserve
our attention and resources in our quest to answer one of modern science's
greatest questions.

\clearpage



\end{document}